\documentclass[aps,prl,superscriptaddress,twocolumn,notitlepage]{revtex4-1}
\usepackage{graphicx}
\usepackage{xcolor}
\usepackage{bm}
\usepackage{amsmath}
\usepackage{amssymb}
\usepackage{amsfonts}
\usepackage{euscript}
\usepackage{verbatim}
\usepackage{setspace}
\usepackage{amsfonts}
\usepackage{braket}
\usepackage{natbib}

\usepackage{array}

\usepackage[normalem]{ulem}
\usepackage{hyperref}

\usepackage[capitalise]{cleveref}

\makeatletter
\newcommand*{\whatisthesize}{\f@size pt}
\makeatother

\newcommand{\be}{\begin{equation}}
\newcommand{\ee}{\end{equation}}
\newcommand{\bea}{\begin{eqnarray}}
\newcommand{\eea}{\end{eqnarray}}

\newcommand*{\yd}{{\dagger}}
\newcommand*{\nd}{{\vphantom{\dagger}}}

\newcommand*{\I}{1\!\!1}

\newcommand{\PreserveBackslash}[1]{\let\temp=\\#1\let\\=\temp}
\newcolumntype{C}[1]{>{\PreserveBackslash\centering}p{#1}}
\newcolumntype{R}[1]{>{\PreserveBackslash\raggedleft}p{#1}}
\newcolumntype{L}[1]{>{\PreserveBackslash\raggedright}p{#1}}

\newcommand{\ev}[1]{{\langle #1\rangle}}

\begin{document}
\title{Dissipative mean-field theory of IBM utility experiment}

\author{Emanuele G. \surname{Dalla Torre}}
\email{emanuele.dalla-torre@biu.ac.il}
\affiliation{Department of Physics, Bar-Ilan University, Ramat Gan 5290002, Israel}
\affiliation{Center for Quantum Entanglement Science and Technology, Bar-Ilan University, Ramat Gan 5290002, Israel}

\author{Mor M. \surname{Roses}}
\affiliation{Department of Physics, Bar-Ilan University, Ramat Gan 5290002, Israel}
\affiliation{Center for Quantum Entanglement Science and Technology, Bar-Ilan University, Ramat Gan 5290002, Israel}

\begin{abstract}
In spite of remarkable recent advances, quantum computers still lack useful applications. A promising direction for such utility is offered by the simulation of the dynamics of many-body quantum systems, which cannot be efficiently computed classically. Recently, IBM used a superconducting quantum computer to simulate a kicked quantum Ising model with large numbers of qubits and time steps. 
{\color{black} These results were later reproduced using numerical techniques based on tensor networks and Clifford expansion. In this work, we analyze the experiment in the eyes of a simple-minded mean-field approximation. We treat neighboring qubits as a self-consistent source of dephasing and express them in terms of Kraus operators. Although our approach completely disregards entanglement between qubits, it captures  the overall dependence of physical observables as a function of time and external magnetic field.
This observation can help rationalize}
the success of the quantum computer in solving this specific problem.
\end{abstract}


\maketitle

{\it Introduction ---}  Quantum computers hold the promise of finding useful applications in combinatorial optimization, quantum chemistry, cryptography decryption, and quantum simulations of many-body quantum systems. In particular, several recent works used quantum computers to study the real-time dynamics of the celebrated transverse-field Ising model in one and two dimensions. In the case of digital quantum computers, the model can be implemented using a Trotter expansion where the transverse field and the Ising coupling act alternatively in discrete steps. Such implementation gives rise to a periodically driven system, whose phase diagram was studied in Ref.~\cite{else2016floquet} and includes ferromagnetic, paramagnetic, time crystal, and Floquet topological phases \footnote{These phases respectively correspond to no Majorana fermion, one Majorana fermion at energy zero, one Majorana fermion at energy $\pi$, and two Majorana fermions.}.  Several works realized this model on quantum computers to study topological edge state \cite{azses2021observing}, confinement \cite{vovrosh2021confinement}, disorder \cite{mi2022time}, Kibble-Zurek scaling \cite{azses2023navigating}, and more \cite{cervera2018exact,zhukov2018algorithmic,oftelie2022computing}.


In a recent work \cite{kim2023evidence}, researchers from IBM studied the dynamics of this model on a 127-qubit quantum computer with up to $t=20$ steps. For $t\leq 5$, the model can be solved exactly using a light-cone and depth-reduced  method, which allows one to effectively reduce the size of the relevant Hilbert space. For longer times, the model can be solved using advanced numerical approximation techniques based on tensor networks \cite{tindall2024efficient,kechedzhi2024effective,patra2024efficient}  and circuit resummation \cite{begusic2023fast}. In both regimes, the experimental results matched quantitatively the exact solutions: Using an error mitigation technique called zero-noise extrapolation \footnote{See Ref.~\cite{kim2023scalable} and references therein.} they were able to obtain results within a few percent error from the exact ones.

This observation is very surprising, considering that the largest circuit includes $2880$ two-qubit gates. Assuming a realistic error per gate of 0.5\% \cite{ibm2022qv}, one obtains that the fidelity of the final state is of the order of $0.995^{2880} \approx 5\times 10^{-7}$. This number is several orders of magnitude smaller than the experimentally observed precision. How come that the present circuit delivered excellent results, in spite of prohibitively large error rates? {\color{black} A possible answer to this question is that there exists a semi-classical description of the circuit that does not rely on superposition between different states and is, therefore, immune to decoherence. The aforementioned numerical techniques do exactly this, by distilling a large (but computationally tractable) number of classical degrees of motion. 

In this work, we follow a different path and aim to derive the semiclassical variables starting from expectation values of physical operators. The method that we use extends upon the common mean-field approach to quantum magnets, by taking into account beyond mean-field fluctuations to lowest order.} Our key finding is that the specific problem studied by the IBM team can be effectively described by the dynamics of a single qubit in the presence of randomly fluctuating fields. This finding {\color{black}can provide some insight into} why the experiment was so successful and noise resilient.

\begin{figure}[t]
 \centering
		\includegraphics[width=\linewidth]{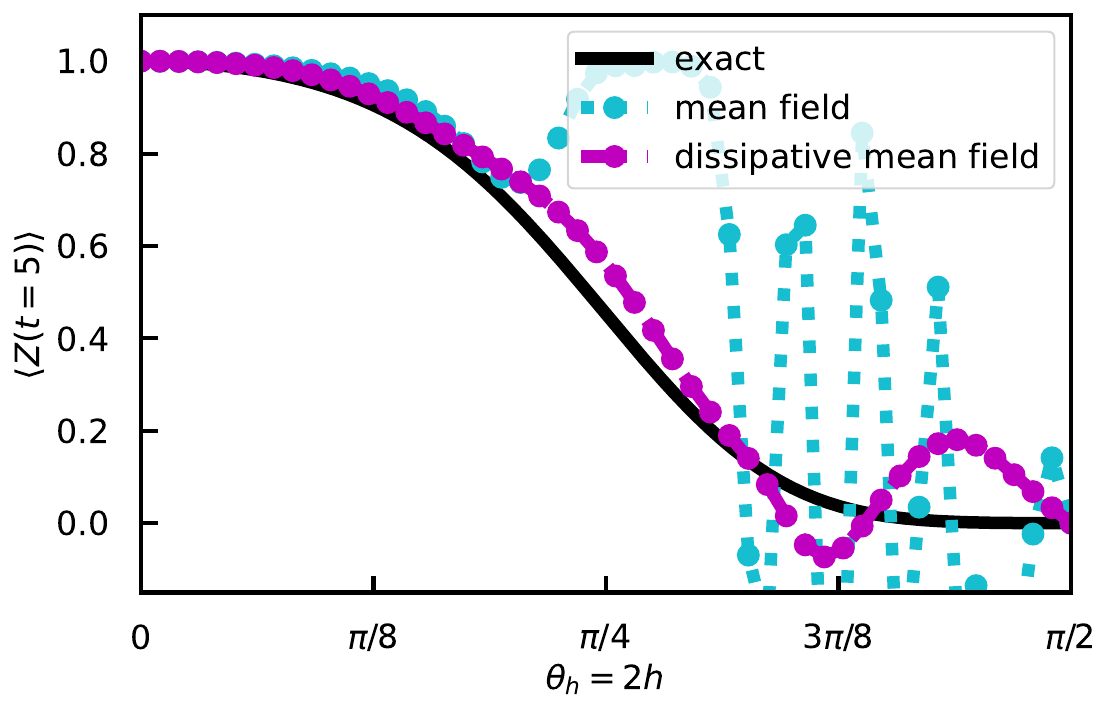}\\
        \includegraphics[width=\linewidth]{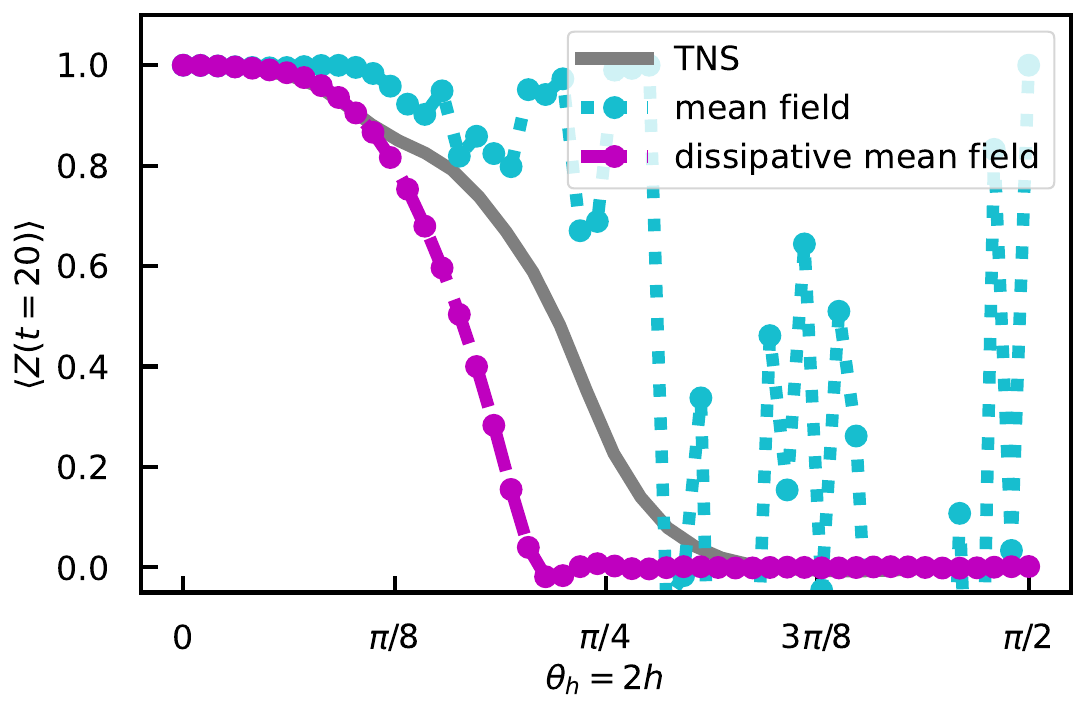}
	\caption{\color{black}Magnetization $\ev{Z}$ after $t=5$ (upper panel) and $t=20$ (lower panel) Trotter steps. The exact solution is the average magnetization reproduced from Ref.~\cite{kim2023evidence} and the tensor-network simulation (TNS) is $\langle Z_{62}\rangle$ reproduced from Ref.~\cite{tindall2024efficient}.  The mean-field approach describes the unitary dynamics of a single qubit affected by a magnetic field determined by the average magnetization of the system. The dissipative mean field approach adds to it an effective dissipative term with a probability given by \cref{eq:p}.}
    \label{fig:fig1}
\end{figure}

The transverse-field Ising model realized by Ref.~\cite{kim2023evidence} is described by the unitary map $U=U_{zz}U_{x}$, where $U_{zz}=\exp( i J \sum_{j,k} \Lambda_{j,k} Z_j Z_k)$ and $U_{x} = \exp(-i h \sum_{j} X_j )$.
Here  $X$ and $Z$ are Pauli matrices,
$i=0,\ldots,L-1$, and $L=127$ is the number of qubits in the device. The connectivity matrix $\Lambda$ is determined by the planar device topology such that $\Lambda_{i,j}=1 (0)$ if the qubits $i$ and $j$ are (not) connected. The device has $M=144$ tunable couplings, such that on average, each qubit is connected to $\xi = \sum_{i,j} \Lambda_{i,j}/L=2M/L \approx 2.27$ neighbors. In the experiment of Ref.~\cite{kim2023evidence} $\theta_J=2J$ was set to $\pi/2$, and $\theta_h=2h$ was varied between 0 and $\pi/2$. See \cref{fig:fig1} for the average magnetization $Z=\sum_i Z_i/L$ after $t=5$ steps as a function of $\theta_h$, computed by exact diagonalization in Ref.~\cite{kim2023evidence}.

To approximate this many-body problem, we first consider a spatially-independent, self-consistent, mean-field theory: we describe the system by a product state where each qubit is found in the same wavefunction $\ket{\psi_0}$. In such mean-field approximation, $\ket{\psi_0}$ evolves according to $u = u_{zz}u_x $, where
\begin{align}
\label{eq:Ux}
    u_x=&\exp(-ih X),\\
    \label{eq:Uzz}
    u_{zz}=&\exp(iJ\xi zZ).
\end{align}
Here, we approximated the magnetization of the neighboring qubits by its average value $z = \ev{Z}$, evaluated at the previous step. The initial condition corresponds to $\ket{\psi_0}=\ket{0}=(1,0)^T$ and the dynamics can be computed numerically by multiplying 2-by-2 matrices. The result of this approach is shown in \cref{fig:fig1}. We find that this simple-minded mean-field approach gives a good approximation for small $h <\pi/8$, while it fails at larger $h$.

To understand the origin of this failure let us consider the limiting case of $\theta_h=\pi/2$. After the first application of $U_x$, all the qubits are rotated around the $X$ axis by an angle of $\pi/2$. This leads to a state where all the qubits point in the $+Y$ direction, $\ket{\psi_0}=\ket{+i}$. Next, they experience an Ising coupling $U_{zz}$. Because the average magnetization is zero, the mean-field approximation gives $u_{zz}=\I$ and has no impact. In contrast, in the full quantum model, the operator $U_{zz}$ has a nontrivial action as it entangles neighboring qubits. Because the state $\ket{+i}$ corresponds to an equal superposition of $\ket{0}$ and $\ket{1}$, when we focus on a single qubit $k$ and integrate out its neighbors, we can effectively treat the latter as random variables with equal probability of being in the 0 and 1 states. Accordingly the operator $U_{zz}$ corresponds to a random rotation of the $j$th qubit and effectively dephases it.

To describe such dephasing process we consider the difference between the actual magnetization of the neighboring qubits and the mean-field value,
\begin{align}
\delta Z_j = \sum_{k} \Lambda_{j,k} Z_k -  \xi z.
\label{eq:deltaZj}
\end{align}
This term affects the dynamics of the $j$th qubits by
\begin{align}
u^\nd_{\delta Z,j} = e^{i J Z_j \delta Z_j }.
\end{align}
The action of this operator on the Pauli matrices $X_j$, $Y_j$ and $Z_j$ is given by
\begin{align}
\label{eq:rotation}
\begin{aligned}
X' \equiv u^\yd_{\delta Z} X u^\nd_{\delta Z} &= X \cos(2J\delta Z) + Y \sin(2J\delta Z),\\
Y' \equiv u^\yd_{\delta Z} Y u^\nd_{\delta Z} &= Y \cos(2J\delta Z) - X \sin(2J\delta Z),\\
Z' \equiv u^\yd_{\delta Z} Z u^\nd_{\delta Z} &= Z,
\end{aligned}
\end{align}
where we dropped the $j$ sub-indexes for brevity. If we consider $\delta Z$ as a stochastic variable, \Cref{eq:rotation} correspond to a random rotation of an angle $2J\delta Z$ around the $Z$ axis. Following a mean-field logic, we assume $\delta Z$ to be a Gaussian variable. This approximation is formally valid in the limit of $\xi\to\infty$, at fixed $J\xi$, such that $\delta Z$ is given by the sum of an infinite number of terms, and the central-limit theorem applies. As we will now show, this approximation gives a reasonable result even for the current problem, where $\xi$ is close to 2 \footnote{See, {\it e.g.}, Ref.~\cite{auerbach2012interacting} for a related discussion about the validity of the large-$N$ limit in the case of $N=2$.}.

Starting from \cref{eq:deltaZj}, it is easy to show that $\delta Z$ has zero average and variance
\begin{align}(\delta z)^2 = \ev{\delta Z^2} = \xi (1-z^2).\end{align}
Here, we used the mean-field expression $\ev{  Z_j  Z_k} = \delta_{i,j} + (1-\delta_{i,j}) z^2$. Using the properties of Gaussian variables, we can average over $\delta Z$ in \cref{eq:rotation} and obtain
\begin{align}
\label{eq:exp}
X' =&  e^{-2( J\delta z)^2} X,&
Y' =& e^{-2( J\delta z)^2)} Y,&
Z' =& Z.
\end{align}
These equations can be combined as
\begin{align}
O' = (1-p) O + p Z O Z,
\label{eq:kraus1}
\end{align}
where $O\in \lbrace X,Y,Z\rbrace$ and $1-2p =\exp(-2( J \delta z)^2)$, or
\begin{align}
p = \frac 12(1-e^{-2( J\delta z)^2}).\label{eq:p}
\end{align}

\Cref{eq:kraus1} describes the evolution of operators in the Heisenberg picture. This process can be equivalently described in the Schroedinger picture by the Kraus map $\rho \to \rho'$, where
\begin{align}
\rho' = (1-p) \rho + p Z \rho Z.
\label{eq:kraus2}
\end{align}
This equation corresponds to a Kraus map with Kraus operators $K_1 = \sqrt{1-p}\I$ and $K_2 = \sqrt{p}Z$. This map is often used to describe single-qubit pure dephasing ($T_\phi$ process) with probability $p$. In this study, the effective dephasing originates from the many-body dynamics, i.e. the $ZZ$ Ising coupling.

In \cref{fig:fig1} we plot the mean-field magnetization $z = \ev{Z} ={\rm Tr}[\rho Z]$, where $\rho$ is computed by applying the unitary operators $u_x$, ${u}_{zz}$, \cref{eq:Ux,eq:Uzz}, and the Kraus map, \cref{eq:kraus2}, sequentially for $t=5$ times. Remarkably, this simple mean-field approach offers a qualitative description without any free fitting parameter. Our theory predicts a small revival in the vicinity of $\theta_h=3\pi/16$, which is not observed in the exact solution. {\color{black}This effect is partially suppressed if one takes into account the inhomogeneity of the connectivity and performs a mean-field dynamics of each qubit separately}. Alternatively, to obtain a better agreement, one can extend the mean-field description to small clusters of qubits \cite{jin2016cluster}. Finally, one can consider retardation effects by introducing the two-time correlation functions $\ev{\delta Z(t)\delta Z(t')}$, giving rise to a dynamical mean-field approximation \footnote{See, {\it e.g.}, Ref.~\cite{cugliandolo2023recent} for a recent review.}. We find it remarkable that the present  approach, which is local in both time and space, is sufficient to capture key aspects of the dynamics of the system.

\begin{figure}[t]
\centering
\includegraphics[width=\linewidth]{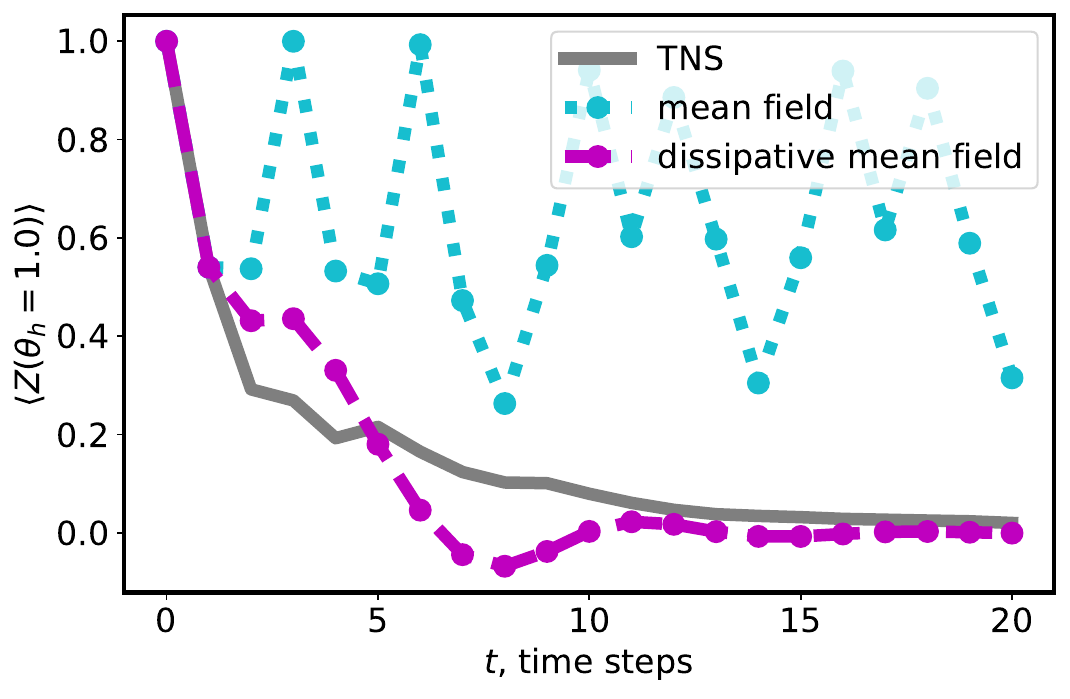}
\caption{\color{black}Same as Fig.~\ref{fig:fig1} as a function of Trotter steps, at a fixed value of $\theta=1.0$.}
\label{fig:time}
\end{figure}

\begin{figure}[t]
\centering
\includegraphics[width=\linewidth]{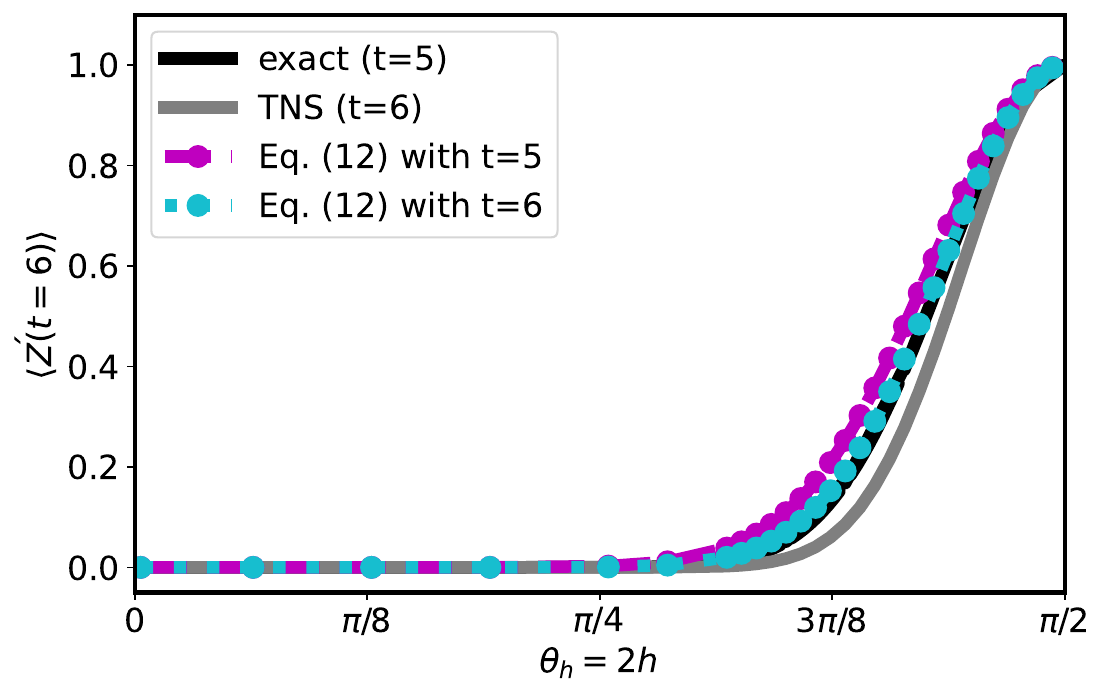}
\caption{\color{black}Expectation value of non-local operators after $t=5$ and $t=6$ steps.}
\label{fig:eff}
\end{figure}

We now turn to the study of non local correlations. In particular, Ref.~\cite{kim2023evidence} considered the expectation value of a stabilizer of the circuit at $\theta_h=\pi/2$. {\color{black} To develop a mean-field theory of this circuit, let us perform a basis transformation, such that at each step, the circuit with $\theta_h=\pi/2$ transforms the $|000...\rangle$ state to itself. In this new basis, which we denote by $Z'$, the circuit at $\theta_h=\pi/2$ is diagonal in the $Z'$. If we move away from $\theta_h=\pi/2$, the circuit will begin to affect the stabilizers by creating some effective coupling between them of the form $\exp(iJ_{j,k}X'_j X'_k)$, $\exp(iJ_{j,k,l}X'_j X'_k X'_l)$ etc. In a stochastic mean-field approach these terms can be treated as random rotations of the form $\exp(i\alpha_j X'_j)$, where $\alpha=\sum_k \Lambda_{j,k} X'_k + \sum_{k,l}\Lambda_{j,k,l} X'_k X'_l + ...$ are random variables. Using the fact that $\langle X'\rangle =0$, we obtain a process of the type described in Eq.~(\ref{eq:kraus2}) with $Z\to X$ and
\begin{align}
p=\frac12\left(1-e^{-2(\delta\alpha)^2}\right),
\end{align}
with $(\delta\alpha)^2 = \langle \alpha^2\rangle$. By construction, $\delta\alpha=0$ at $\theta_h=\pi/2$ and obtains its maximal value $\delta\alpha=\pi/2$  at $\theta_h=0$. If we assume a linear dependence, we arrive at the expression: 
\begin{align}
(\delta\alpha)^2 = \frac{\pi^2}{4}\left(1-\frac{2\theta_h}\pi\right)
\end{align}
Because $\langle X'\rangle=0$, the evolution of $\langle Z'(t)\rangle$ can be simply described as stochastic process that starts with at 1 and flips its sign with probability $p_0$ at each step. After $t$ steps one obtains
\begin{align}
\ev{Z'(t)} = (1-2p_0)^t = \exp\left[-t\frac{\pi^2}2\left(1-\frac{2\theta_h}\pi\right)^2\right].\label{eq:eff}
\end{align}}
As shown in \cref{fig:eff}, this analytic expression captures the main features of the numerically exact results.

To summarize, in this Letter, we addressed the surprising precision of noisy quantum computers in simulating the dynamics of the kicked quantum Ising model, despite the seemingly prohibitive accumulation of errors in the process. To understand this effect, we introduced a dissipative mean-field theory that gives a possible explanation to this puzzle. Within this approximation, {\color{black} the evolution of the expectation values probed in the experiment} is equivalent to the much simpler dynamics of a single qubit undergoing rotations and dephasing. In such a model, the natural decay and dephasing of the physical qubits have little effect, as long as these processes are slower than those induced by the many-body circuit. On general ground, physical observables that can be efficiently described using mean-field theories that are local in both time and space, such as the present one, are easier to realize on noisy quantum computers. Unfortunately, this is also the regime where classical simulations are very efficient and no utility of quantum computers is expected. Conversely, our work can guide experimentalists towards observable that are hard to simulate classically, where quantum computers may become useful in the future.

\begin{acknowledgments}
{\bf Acknowledgments} We acknowledge useful discussions with David Dentelski. This work is supported by the Israel Science Foundation, grant number 151/19, 154/19, and 2126/24. This research was initiated at the International Centre for Theoretical Sciences (ICTS) during the Periodically and quasi-periodically driven complex systems workshop (code:  ICTS/pdcs2023/6) and advanced at the Kavli Institute for Theoretical Physics during the 
Quantum Optics of Correlated Electron Systems program (NSF PHY-2309135). An appendix to the original version of the submitted manuscript has now been published online as a separate work in Ref.~\cite{practitioner2025dalla}.
\end{acknowledgments}

\bibliography{references}

\end{document}